\documentclass[11pt]{article}

\usepackage{cite}
\usepackage{amsmath, amsfonts}
\usepackage{booktabs}
\usepackage[english]{babel}
\usepackage{graphicx}
\usepackage{multirow}
\usepackage{subfig}
\usepackage{hyperref}
\usepackage[range-units=brackets, range-phrase=-, per-mode=reciprocal, mode=math]{siunitx}
\usepackage[babel=true]{microtype}

\numberwithin{equation}{section}
\allowdisplaybreaks[1]
\def\beq{\begin{equation}}
\def\eeq{\end{equation}}

\usepackage{colortbl}

\makeatletter
\newlength{\apb@width}
\newcommand{\autoparbox}[2][c]{\settowidth{\apb@width}{#2}\parbox[#1]{\apb@width}{#2}}

\makeatother

\definecolor{lightgray}{gray}{0.9}

\usepackage[framemethod=default]{mdframed}
\newmdenv[skipabove=7pt,
skipbelow=7pt,
rightline=false,
leftline=false,
topline=false,
bottomline=false,
backgroundcolor=gray!10,
linecolor=gray,
innerleftmargin=5pt,
innerrightmargin=5pt,
innertopmargin=5pt,
innerbottommargin=5pt,
leftmargin=0cm,
rightmargin=0cm,
linewidth=4pt]{eBox}

\numberwithin{equation}{section}

\def\bea{\begin{eqnarray}}
\def\eea{\end{eqnarray}}

\def\beq{\begin{equation}}
\def\eeq{\end{equation}}
\def\bea{\begin{eqnarray}}
\def\eea{\end{eqnarray}}

\def\k{{\bf k}}

\def\p{{\bf p}}
\def\x{{\bf x}}

\DeclareRobustCommand{\SkipTocEntry}[4]{}

\definecolor{blue3}{RGB}{31, 119, 180}
\definecolor{red3}{RGB}{	214, 39, 40}
\definecolor{orange3}{RGB}{255, 127, 14}
\definecolor{green3}{RGB}{44, 160, 44}

\begin{document}

\begin{titlepage}
\setcounter{page}{1} \baselineskip=15.5pt 
\thispagestyle{empty}

\begin{center}
{\fontsize{18}{18} {\bf Signals of a Quantum Universe} }
\end{center}

\vskip 20pt
\begin{center}
\noindent
{\fontsize{12}{18}\selectfont  Daniel Green$^1$ and Rafael A. Porto$^2$}
\end{center}

\begin{center}
\vskip 4pt
\textit{$^1$ Department of Physics, University of California, San Diego, La Jolla, CA 92093, USA}
\vskip 8pt
\textit{$^2$ Deutsches Elektronen-Synchrotron DESY,
Notkestrasse 85, 22607 Hamburg, Germany}
\end{center}

\vspace{0.4cm}
 \begin{center}{\bf Abstract}
 \end{center}
 
 \noindent
Structure in the Universe is widely believed to have originated from {\it quantum} fluctuations during an early epoch of accelerated expansion. Yet, the patterns we observe today do not distinguish between quantum or classical primordial fluctuations; current cosmological data is consistent with either possibility. 
We argue here that a detection of primordial non-Gaussianity
 can resolve the present situation, and provide a litmus-test for the quantum origin of cosmic structure. 
Unlike in quantum mechanics, vacuum fluctuations cannot arise in classical theories and therefore long-range classical correlations must result from 
(real) particles in the initial state. 
Similarly to flat-space scattering processes, we show how basic principles require these particles to manifest themselves as poles in the $n$-point functions, in the so-called folded configurations.
   Following this observation, and assuming fluctuations are 
    {\it(i)} correlated over large scales, and {\it (ii)} generated by local evolution during an inflationary phase, we demonstrate that: {\it the absence of a pole in the folded limit of non-Gaussian correlators 
    uniquely identifies the quantum vacuum as the initial state}. In the same spirit as Bell's inequalities, we discuss how this can be circumvented if locality is abandoned. We also briefly discuss the implications for simulations of a non-Gaussian~universe.

\end{titlepage}

\newpage
\setcounter{tocdepth}{2}

  \makebox[0pt][l]{%
  \raisebox{-0.934\totalheight}[0pt][0pt]{%
\hskip -18pt
    \includegraphics[width=2.75in]{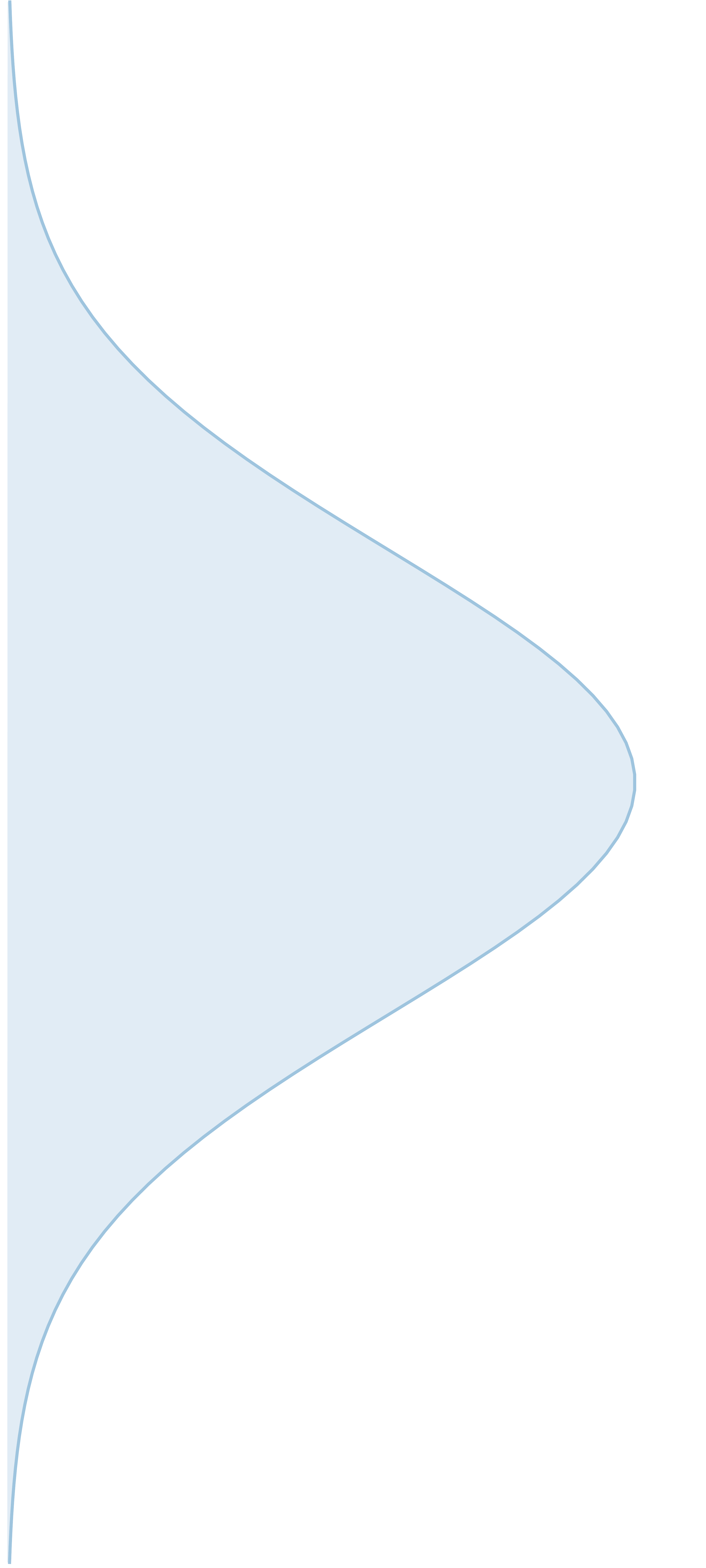}}}%
\vskip 3cm
\tableofcontents

\newpage

\section{Introduction}

Cosmological observations strongly suggest that structure in the universe originated from minute fluctuations present in the very early universe, prior to the hot big bang~\cite{Hu:1996yt,Spergel:1997vq,Dodelson:2003ip}.  A compelling possibility is that these density perturbations were produced through quantum mechanical zero-point fluctuations in the vacuum~\cite{Mukhanov:1981xt,Hawking:1982cz,Guth:1982ec,Starobinsky:1982ee,Bardeen:1983qw}, and then were stretched over long distances by rapid accelerated expansion (inflation). In one brush, this idea unveils a beautiful connection between the largest structures in the cosmos and the fundamental laws of physics at the smallest scales. Yet, current data~\cite{Akrami:2018odb,Akrami:2019izv} could equally be explained if inflation had stretched classical statistical fluctuations instead. In the same fashion as Bell's program back in the 1960's put quantum mechanics to the test~\cite{Bell:1964kc}, our goal here is to bring the quantum origin of the density fluctuations, realized in a majority of models, into a well-defined statement that can be confronted with future observations.

Unfortunately, one cannot simply perform experiments with the entire universe. We only get to observe the one we inhabit, and only have access to an effectively classical probability distribution of fluctuations~\cite{Grishchuk:1990bj}.  Classic tests of quantum mechanics, such as Bell's inequalities~\cite{Bell:1964kc}, cannot be directly applied in this case. As a result, despite a long history (e.g.~\cite{Starobinsky:1986fx,Grishchuk:1990bj,Campo:2005sv,Lim:2014uea,Martin:2015qta,Goldstein:2015mha,Nelson:2016kjm,Choudhury:2016cso,Martin:2017zxs,Shandera:2017qkg,dePutter:2019xxv}), there has been limited progress identifying observational connections between the quantum initial state and the classical universe we observe today. Recently, a step towards a potential signature was suggested by Maldacena~\cite{Maldacena:2015bha}. For a judiciously chosen model, the dynamics during inflation effectively performs a Bell-type measurement, storing the result in the final probability distribution. The proposal does not suggest a generic observational test; yet, although baroque, Maldacena's model is a proof of principle that the primordial fluctuations can {\it remember} their quantum origin.

\begin{figure}[!t]
\includegraphics[width=\textwidth]{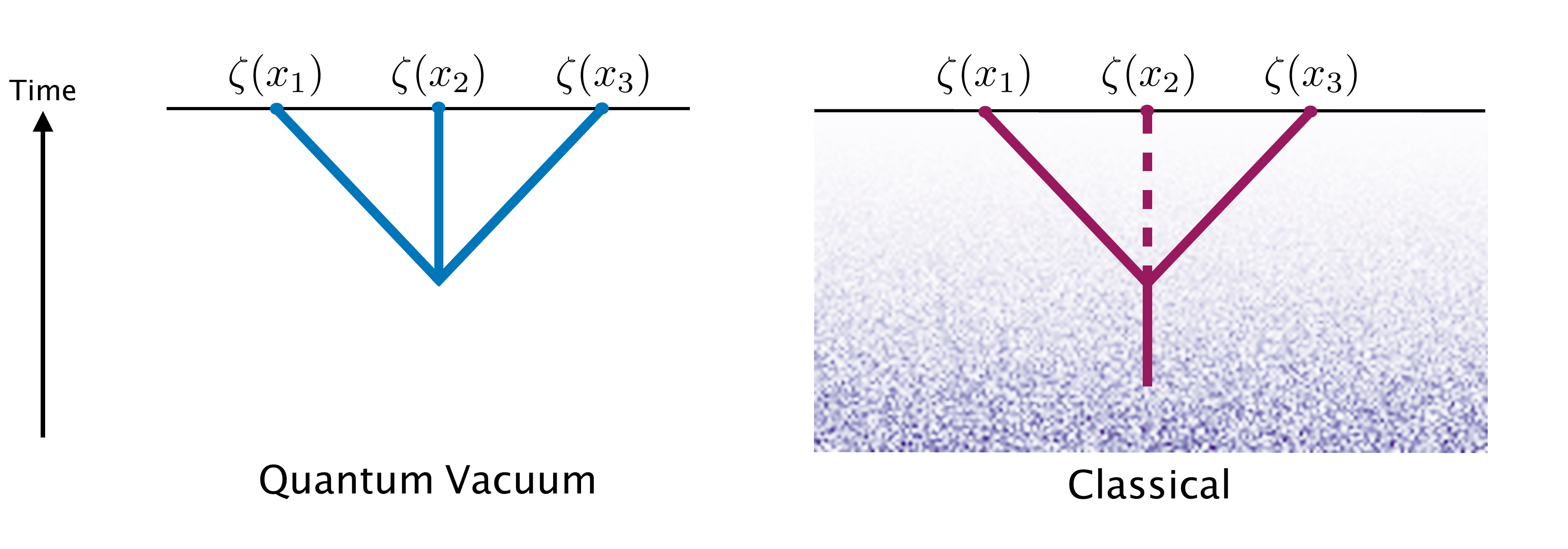} 
      \caption{Late-time observations measure correlations of the adiabatic density fluctuation, $\zeta(\x,\tau)$, produced from non-linear time evolution in the early universe. The particle's propagation is illustrated by the solid lines, while the dashed line represents the absence of the corresponding mode at late times. {\it Left:} Quantum-vacuum fluctuations arise as the correlated production of three particles due to non-linear effects. This process would violate energy conservation in flat space, and thus produces no poles at physical momenta~\cite{Flauger:2013hra}. {\it Right:} Classical fluctuations only arise in a state containing physical particles, as local variations in the particle density, e.g.~\cite{Berera:1995ie,Berera:1998px,Green:2009ds,LopezNacir:2011kk,LopezNacir:2012rm,Turiaci:2013dka}.  The non-linear evolution that leads to net particle creation also allows for decays (or annihilation). These processes appear as poles at physical momenta.}
      \label{fig1}
\end{figure}

In this letter we will pursue these ideas further, and provide a testable prediction of the quantum nature of the initial state. We will argue that non-linear local evolution of the density fluctuations can indeed store its quantum origin in the correlations observed at late times.  
Concretely, we will show how only quantum mechanics can produce the type of long-range correlations typical of the vacuum state, while classical fluctuations are necessarily produced by (highly-excited) states with their own characteristic features.
The basic picture, illustrated in Fig.~\ref{fig1}, is the following: Non-Gaussian correlations in the quantum-vacuum are associated with `particle-creation'. In~contrast, and due to locality, causal classical evolution must also include the decay of particles in the initial state. Hence, even though both vacuum and classical effects produce correlations on large scales at late times, the latter necessarily encode its distinctive physical origin, yielding distinguishable signatures from the case of quantum-vacuum fluctuations. In particular --- in analogy with flat-space {\it polology}~\cite{Weinberg:1995mt} --- an associated pole must be present for classical $n$-point functions (beyond the power spectrum). Moreover, a {\it width} will also be generated,  
through dissipation~\cite{Berera:1995ie,Berera:1998px,Green:2009ds,LopezNacir:2011kk,LopezNacir:2012rm,Turiaci:2013dka}, which effectively smooths these poles to produce a {\it bump} at physical momenta, as in particle colliders.

The existence of poles by itself may not be sufficient to show that classical physics is the culprit. For instance, quantum excited states  can also develop the same pole structure~\cite{Flauger:2013hra}. Yet, we will demonstrate that {\it the absence of this signature --- in otherwise observable long-range non-Gaussian correlations --- can only be explained by quantum zero-point effects.}  
In~other words, in a classical framework consistent with locality, tampering with the analytic structure of the correlators in an attempt to remove the poles, will unavoidably alter the structure at large scales, as expected from our intuition in flat space. On the other hand, long-range correlations --- as those featured in the vacuum state --- may be produced without the associated poles if locality is violated. We will illustrate the role of local causal evolution in an illuminating example.

Our analysis is also motivated by the practical issue of simulating a universe with non-Gaussian initial conditions.  Typically, generating initial conditions with non-local correlations from a Gaussian map requires high-dimensional integration~\cite{Smith:2006ud,Schmidt:2010gw,Scoccimarro:2011pz}. If these initial conditions were generated by local classical evolution instead, one could simply produce them via a Gaussian map evolved in time, and potentially speed up the simulations.\footnote{We thank Uro\v s Seljak for first raising this point.}
However, as we show here, such a procedure --- or any local evolution for that matter --- will not accurately reproduce the non-Gaussian probability distribution obtained from quantum fluctuations. This result may also have some deeper relevance in quantum versus classical computing.

\section{Cosmic Quantumness}

We are interested in comparing the predictions of quantum and classical physics for the statistics of the initial density perturbations, assuming that the non-Gaussianity is produced from local non-linear evolution, namely it is not present in the initial state. 

\subsection*{Gaussian Fluctuations} 

For concreteness, we will assume that the adiabatic density fluctuations, $\zeta(\x,\tau)$, arise from an effectively massless field during inflation propagating in a de Sitter background,\footnote{For illustrative purposes, we will ignore slow-roll corrections, e.g. the spectral tilt.}  
\beq
ds^2 =  -dt^2 + a(t)^2 d\x^2 = a(\tau)^2 (-d\tau^2 + d \x^2) = \frac{1}{H^2\tau^2} (-d\tau^2 + d \x^2) \,,
\eeq
with $a$ the scale factor in physical $(t)$ and conformal $(\tau)$ time, respectively. Recall the (constant) Hubble expansion parameter is given by $H \equiv \dot a(t)/a(t)$, where we use the notation (throughout this paper) $\dot f \equiv \partial_t f = a^{-1}\partial_\tau f$, for derivatives w.r.t. the physical time.

The modes of the density perturbations obey
\beq
\zeta(\x,\tau) = \int \frac{d^3 k}{(2\pi)^3} \frac{\Delta_\zeta} {\sqrt{k^3}} e^{i \k\cdot \x} [ a^\dagger_\k (1-i k\tau) e^{i k \tau} +a_{-\k} (1+i k\tau) e^{-i k \tau}  ] \ , \label{eq:norm}
\eeq
where the normalization, $\Delta_\zeta$, is so chosen to coincide with the observed amplitude of adiabatic fluctuations.
Since the field is real, we have $(a^\dagger_\k)^\dagger= a_{-\k}$. The statistical differences arise once we compare quantum versus classical correlation functions. 
\begin{itemize}

 \item {\bf Quantum}. The $a^\dagger_\k$ are creation {\it operators} in a Hilbert space, satisfying:
 \beq
 [a^\dagger_\k, a_{\k'}] = \delta(\k - \k') \qquad a_\k |0 \rangle = 0\,,
  \eeq
which readily imply 
  \beq
  \langle 0|  a_{\k'} a^{\dagger}_\k |0 \rangle = \delta(\k-\k')\,, \qquad \langle 0| a^{\dagger}_\k a_{\k'}|0 \rangle = 0 \,,
  \eeq
  in the vacuum state.  In what follows, we will define $\langle 0| [\ldots]|0\rangle \to \langle [\ldots]\rangle_q$ for  convenience.
  \item {\bf Classical}. The $a^\dagger_\k$ are stochastic {\it parameters}, which obey the following statistical properties:
\beq
\langle a^{\dagger}_\k a_{\k'}\rangle_c =\frac{1}{2}  \delta(\k-\k') = \langle  a_{\k'} a^{\dagger}_\k \rangle_c\,,
 \eeq
 as an ensemble average. Notice that the second equality is only valid for classical fluctuations, since it implies that the stochastic parameters commute.  
 \end{itemize}
Although produced by different mechanisms, both the classical and quantum-vacuum fluctuations are normalized to give rise to the same equal-time correlation function,
\beq
\langle \zeta_\k(\tau) \zeta_{\k'}(\tau) \rangle = \frac{\Delta_\zeta^2}{k^3} (1+k^2 \tau^2) \, (2 \pi)^3 \delta(\k+\k') \,,
\eeq
in the absence of interactions. Therefore, measurements of the power spectrum alone are not sufficient to distinguish between them. On the other hand, for unequal times the quantum and classical two-point functions do not agree, reflecting the non-zero commutator, $[\zeta(\x,\tau),\dot \zeta(\x,\tau)]\neq 0$, in the quantum theory. This distinction plays a key role when interactions are present.

We will illustrate the main difference explicitly in the next section. However, we can also understand the root of the discrepancy as follows. Let us introduce the real and imaginary parts of $\zeta$, such that 
 \beq\label{eq:sines}
\zeta(\x,\tau) = \int \frac{d^3 k}{(2\pi)^3} \frac{\Delta_\zeta} {\sqrt{k^3}} e^{i \k\cdot \x} \left[ a_{R,\k} \left(\cos(k \tau)+k \tau \sin(k \tau) \right) + a_{I,\k} \left(\sin(k \tau)  - k \tau \cos( k \tau) \right) \right]  \ , 
\eeq
with $a_\k = \frac{1}{2} [a_{R,\k} + i a_{I,\k}]$. In these variables, the classical statistics obey
\beq \langle a_{R, \k} a_{R,\k'} \rangle_c = \delta(\k+\k')  = \langle a_{I, \k} a_{I,\k'} \rangle_c\, ,
\eeq 
which implies that the independent fluctuations are sines and cosines.  
On the other hand, in relativistic quantum mechanics --- due to locality/causality ---  we speak instead of positive {\it and} negative frequency (energy) modes. The former are {\it annihilated} (by definition) in the vacuum state, while virtual particles can be produced. This textbook observation is the key that allows us to demonstrate how non-linear interactions can discern between classical and quantum correlations.

\subsection*{Non-Gaussianity}
In order to gain intuition, we will consider an illustrative example with the interaction Hamiltonian $H_{\rm int} = -\frac{\lambda}{3!} \dot \zeta^3$.  This choice will allow us to perform explicit computations without losing generality.\footnote{We assume cubic interactions are not forbidden by symmetries. (Notice, though feeble, gravity unavoidably sets a~floor for the bispectrum~\cite{Maldacena2,Cabass:2016cgp}.) It is straightforward to show the same result applies to all the $n$-point~functions.} As we shall see, our conclusions will be rooted in well-established principles, and therefore do not depend on the type of interaction as long as it respects locality (see Appendix~\ref{app:comm}). 
\vskip 10pt
\noindent {\bf Quantum}.  The standard (in-in) calculation~\cite{Weinberg:2005vy} in the vacuum state yields~(with $|\k_i|=k_i$)
 \bea
  \label{vacuum} 
 \langle \zeta_{\k_1} \zeta_{\k_2} \zeta_{\k_3} \rangle'_q &=& i\int d\tau'\langle[ H_{\rm int}(\tau'), \zeta_{\k_1} \zeta_{\k_2} \zeta_{\k_3}(0)] \rangle  \\
 &=&  2 \frac{\lambda}{H
 }  \frac{\Delta_\zeta^6}{k_1 k_2 k_3}\, {\rm Im} \int_{-\infty}^0 d\tau' \tau'{}^2 e^{i (k_1+k_2+k_3)\tau'} \nonumber \\
 &=&  \frac{4 \lambda H^{-1} \Delta_\zeta^6}{ (k_1+k_2+k_3)^3 k_1 k_2 k_3}  \ ,\nonumber
 \eea
up to the momentum conserving $\delta$-function, which is denoted by the primed brackets $\langle \rangle'$. Notice, for $k_i\neq 0$, we have a pole in the total energy: $k_t \equiv k_1+k_2+k_3$. This is due to the fact that, for cosmological (in-in) correlators, the {\it would-be} energy-conserving $\delta$-function becomes a factor of $1/k_t^n$, for non-negative integer $n$. Via analytic continuation, as  $k_t \to 0$, the residue of this pole is intimately connected to the flat-space $S$-matrix, with the order of the pole ($n=3$ in this case) related to the number of derivatives at the local interaction.  
In the quantum vacuum, the correlation is produced by the creation of three (virtual) particles ($0 \to 3$), which are subsequently {\it measured} at later times (see Fig.~\ref{fig1}). The uncertainty principle in an expanding universe permits a --- minimal amount of --- violation of  the conservation of energy, $\Delta t \sim H^{-1}$, which is forbidden~classically.  As expected, since there are no real particles to scatter in the vacuum, there are no other processes allowed nor poles at physical momenta. 
\vskip 10pt
\noindent {\bf Classical}. We determine the bispectrum by solving the classical equations of motion perturbatively.  Using the Green's function, obeying
\beq\label{eq:Green2}
 \partial_{\tau'}G_\k(\tau, \tau') = 2 \Delta_\zeta^2  \left[k^{-1}\tau' \sin(k (\tau-\tau')) -  \tau \tau' \cos( k(\tau-\tau'))\right] ,
\eeq
we find at first order in $\lambda$
\beq\label{eq:classicalevol}
\zeta^{(2)}_\k(\tau) = \lambda \int \frac{d\tau' d^3 p }{(2\pi)^3} (-H\tau')^{-1} \left( \partial_{\tau'} G_\k(\tau,\tau') \right) \partial_{\tau'}\zeta^{(1)}_\p(\tau')  \, \partial_{\tau'}\zeta^{(1)}_{\k -\p} (\tau') \ ,
\eeq
where $\zeta_\k^{(1)}$ represents the Gaussian field. Hence, using $\zeta_\k \approx \zeta_\k^{(1)}+\zeta_\k^{(2)}$, the leading contribution to the bispectrum at $\tau = 0$ becomes
\beq
\begin{aligned}
\langle \zeta_{\k_1} \zeta_{\k_2} \zeta_{\k_3} \rangle'_c = \frac{ \lambda  H^{-1}   \Delta_\zeta^6}{6 k_1 k_2 k_3 } \bigg[ \frac{1}{k_t^3}+ \frac{1}{(k_1 +k_2-k_3)^3} &+ \frac{1}{(k_1 -k_2+k_3)^3} \\
\hskip 6cm &+ \frac{1}{(k_1 -k_2-k_3)^3} +\text{permutations} \bigg]\\
= \frac{ \lambda  H^{-1}   \Delta_\zeta^6}{3 k_1 k_2 k_3 } \bigg[ \frac{3}{k_t^3}+ \frac{1}{(k_1 +k_2-k_3)^3} &+ \frac{1}{(k_1 -k_2+k_3)^3}+ \frac{1}{(k_2 -k_1+k_3)^3}\bigg] \label{classic}  \,.
\end{aligned}
\eeq
As anticipated, there are poles at physical momenta in addition to the one at $k_t=0$. These poles are due to classical fluctuations of physical (real) particles in the initial state, which (non-linearly) interact to produce long-range non-Gaussian correlations (see Fig.~\ref{fig1}). For instance, physical particles can decay (annihilate) via on-shell $1 \to 2$ ($2\to1$) processes, and therefore are associated with the poles in the so-called {\it folded limit}~\cite{Babich:2004gb}, where $k_1 \to k_2 +k_3$ and permutations thereof. 
\vskip 10pt
\noindent {\bf Signatures of Quantum Origin:}. The above example illustrates a general property of (in-in) inflationary correlation functions: poles at physical momenta arise from the scattering of real particles present in the initial state. For quantum-vacuum fluctuations there are no real particles, only virtual, yet the poles are still present (by analytic continuation) at negative energies. This is more than just an isolated result mimicking our flat-space intuition. In fact, notice that the overall coefficients of the poles, either in the quantum (\ref{vacuum}) or classical correlation (\ref{classic}) are related, and ultimately linked to the scattering amplitude in the flat-space limit~\cite{Maldacena:2011nz,Raju:2012zr,Arkani-Hamed:2015bza, Arkani-Hamed:2018kmz,Arkani-Hamed:2018bjr,Benincasa:2018ssx}.  Hence, following basic principles, causality guarantees that any process that creates (real) particles at local events, is necessarily accompanied by physical poles in the correlation functions~\cite{Lehmann:1954rq}. The specific form of the interaction controls the resulting polynomial in momentum and/or time dependence, and hence only affects the residue of the poles.\footnote{The overall cancellation in vacuum of the (intermediate) poles arising at physical momenta can be also seen as the appearance of negative residues, due to virtual processes. In fact, the structure of the poles is a direct consequence of the presence of both positive {\it and} negative frequency modes in the classical computation, whereas the vacuum annihilates the former. For classical (highly-excited) states, there is a mismatch between positive and negative contributions, due to real particles in the initial state, leading to the remaining poles seen in \eqref{classic}.} Let us emphasize that this is an unavoidable conclusion, which does not depend on the form of the (local) interaction.  
As a consequence, since there are no vacuum fluctuations in classical mechanics, quantum mechanics is the only way we can guarantee a non-Gaussian signal without violations of locality/causality, while avoiding the existence of poles at physical momenta.

As usual~\cite{Weinberg:1995mt}, decay processes will introduce a finite width which softens the behavior in the folded limit. However, unlike the drift towards the complex plane found in flat space, for (in-in) correlators in an expanding universe the poles move away from the `mass-shell' but remain real. 
While the existence of a width usually happens at higher orders in perturbation theory, models with strong dissipation will exhibit this softening already at tree-level~\cite{LopezNacir:2011kk,LopezNacir:2012rm}, see Appendix~\ref{app:diss}.

\section{Classical Non-localities}

A crucial aspect of Bell's inequalities is that they may be circumvented by non-local theories with hidden-variables at the classical level~\cite{Bell:1964kc,Bohm:1951xw,Bohm:1951xx}. Similarly, locality plays a key role in inferring the quantum nature of the cosmological signal. For our purposes here, it will be sufficient to find an example of a theory which reproduces the same correlators as in the quantum vacuum, but violates~locality. At the same time, we will show that enforcing local causal evolution, while attempting to remove the poles, also alters the type of long-range correlations that are expected in the vacuum state.
\subsection*{Hidden-variables}
For illustrative purposes, we consider a {\it complex} scalar field, which may be decomposed as
\beq
\phi_\k (\tau) = \frac{\Delta_\phi}{k^{3/2} } [ a^{\dagger}_{\k} (1-i k \tau) e^{i k \tau}+ b_{-\k} (1+i k \tau) e^{-i k \tau}]\,,
\eeq
obeying classical Gaussian statistics,
\beq\label{eq:complex}
\langle a^{\dagger}_\k a_{\k'} \rangle_c = \langle  a_{\k'} a^{\dagger}_\k \rangle_c = \delta(\k-\k')\,, \qquad \langle b_\k b^{\dagger}_{\k'} \rangle_c = 0\,. 
\eeq
Let us assume now the existence of a Lagrangian such that the following modified Green's function
\beq
\partial_{\tau'}G_\k(\tau \to 0, \tau')  \to G^{\rm eff}_{\k}(\tau \to 0,\tau') = \frac{1}{k}e^{- i k \tau'} \,,\label{green2}
\eeq
applies in the $\tau \to 0$ limit. Notice that it includes only positive frequency modes. We will not specify the nature of the interaction leading to the above properties, and therefore we treat it as a hidden-variable theory. Using \eqref{eq:classicalevol} with $\zeta \to \phi$  and $\lambda \to \lambda_\phi$, we find 
\beq\label{eq:nonlocal}
\phi_\k(\tau \to 0) =  \frac{i}{3} \lambda_\phi \int \frac{d^3 p}{(2\pi)^3} \int d\tau' \frac{1}{k}e^{- i k \tau'} \dot \phi^*_{\p}(\tau') \dot \phi^*_{\k-\p}(\tau')\,.
\eeq
From this expression we can calculate the bispectrum as usual, yielding
\bea\label{eq:nonlocal_bi}
\langle \phi_{\k_1}\phi_{\k_2} \phi_{\k_3} \rangle' &=& i \lambda_\phi H^{-1} \frac{\Delta_\phi^6}{k_1 k_2 k_3} \int_{-\infty}^0 d\tau' \tau'{}^2 e^{-i (k_1+k_2+k_3)\tau'}  \nonumber \\
 &=& \frac{2 \lambda_\phi H^{-1} \Delta_\phi^6}{(k_1+k_2+k_3)^3 k_1 k_2 k_3}\,. 
\eea
Since $\phi$ is real at late times, it can be converted into density fluctuations after inflation, e.g. $\phi(\tau \to 0 )\approx \zeta$. Up to an overall constant, the result reproduces the same statistical map for the quantum vacuum in \eqref{vacuum} (up to higher order effects which are not relevant here). This theory, however, is non-local as we can see directly from~\eqref{green2}. In particular, 
locality demands that the Green's function in coordinate space must vanish outside of the light-cone. Yet, we have $G^{\rm eff}_{\k}(\tau' \to 0) \simeq k^{-1}$, resembling the Coulomb potential, which is non-zero everywhere in space. As~a consequence, this theory propagates information instantaneously everywhere in space-time.

\subsection*{Causality}

The failure in the above example is rooted in basic principles. Causality in a relativistic theory demands the presence of a negative frequency mode (`anti-particle')~\cite{feynman_feynman_weinberg_1987},\footnote{\footnotesize\url{https://www.youtube.com/watch?v=MDZaM-Bi-kI}} which is precisely what gives rise to the poles at physical momenta.\footnote{We can see this connection more explicitly as follows: The Gaussian theory preserves an effective $U(1)$ charge and~(\ref{eq:complex}) corresponds to an excitation of particle anti-particle pairs.  The non-linear evolution in~(\ref{eq:nonlocal}) breaks this charge conservation and allows three anti-particles to annihilate into nothing (energy need not be conserved in the cosmological background).  The correlation we observe in~(\ref{eq:nonlocal_bi}) is the net gain of 3 particles due to this annihilation.} 
Local causal evolution requires that the Green's function (effectively) must take the form
\beq\label{eq:local_Green}
G^{\rm eff,causal}_{\k}(\tau\to 0, \tau') \propto \frac{1}{k} \sin(k \tau')\,, 
\eeq 
as seen in~(\ref{eq:Green2}), including the negative frequency modes. Unlike three anti-particles annihilating into nothing, as in the non-local example, a causal theory --- with a non-zero number of particles in the initial state --- must also produce poles at physical momentum (in the folded limit), from particle-creation and annihilation at local events. 

In principle, the reader may object that causality could allow for a Green's function which is analytic everywhere in $\k$, even at finite $\tau'$, e.g. 
\beq G^{\rm eff,causal}_{\k}(\tau\to 0, \tau') = \tau'^{\alpha} k^{2\beta}\,,\label{template2}\eeq
where $\beta$ is a non-negative integer. By construction, causality is preserved without positive and negative frequency modes. However, because of analyticity, the Green's function is localized in real space. Hence, while it always vanishes outside the light-cone, it also does not generate long-range correlations.  Such a Green's function can only arise in the limit in which the particle's velocities are small. This is, of course, consistent with the absence of anti-particles in non-relativistic theories. However, this theory must be `UV-completed' into a (local) relativistic one, which will then re-introduce positive and negative frequency modes. More generally, Green's functions with non-analytic behavior in $\k$, necessarily require cancellations between positive and negative frequency modes to remain causal in the $\tau'\to 0$ limit. This is precisely the case in our example, see e.g.~(\ref{eq:local_Green}). (Amusingly, using templates of the form in \eqref{template2}, the associated bispectra takes a form similar to the ones used in computationally efficient simulations~\cite{Scoccimarro:2011pz}.)

The analytic properties of the correlators can also be modified due to the presence of a finite width. In that case, the positive and negative frequency modes will appear in the Green's function, but the would-be divergence is regulated in the folded limit. Hence, the poles are moved slightly away from their physical `on-shell' values. Similarly to the non-relativistic case, a finite width can also alter the type of long-range correlations in otherwise local/causal theories. We elaborate on this point in Appendix~\ref{app:diss}.

\section{Conclusions and Outlook}

The origin of structure as a result of vacuum fluctuations is a purely quantum mechanical phenomenon, for classical effects can only arise when states contain (many) physical particles. Moreover, due to causality, non-linear interactions that allow for the creation of particles must be accompanied by processes in which particles are also able to decay. While the creation of (virtual) particles is allowed, decays are forbidden in vacuum, which gives rise to a dramatic difference in the types of non-Gaussian correlations arising in classical versus quantum-vacuum fluctuations.

The distinction between the two results, as well as the role of locality, is also manifest in the manipulations involved in the derivation of the three-point function. In general, one can show that the difference between the quantum-vacuum and classical computation may be written as:
\begin{align}
 &\qquad \qquad \qquad \qquad\qquad\langle \zeta(\x_1,\tau)\zeta(\x_2,\tau) \zeta(\x_3,\tau)\rangle_q -\langle \zeta(\x_1,\tau)\zeta(\x_2,\tau) \zeta(\x_3,\tau)\rangle_c = \\
 & \frac{i \lambda}{24}  \sum_\sigma \int_{-\infty}^{\tau} d^3 \x' d\tau' a^4(\tau') [\zeta(\x_1, \tau),\hat D_{\sigma(1)} \zeta(\x', \tau')][\zeta(\x_2, \tau),\hat D_{\sigma(2)} \zeta(\x', \tau')][\zeta(\x_3, \tau),\hat D_{\sigma(3)} \zeta(\x', \tau')]\,, \nonumber
\end{align}
where $\hat D_{\ell=1,2,3}$ are local differential operators that characterize the type of interaction(s), and $\sigma$ is a permutation (see Appendix~\ref{app:comm}). The above expression neatly illustrates the link between late time measurements in a quantum state and Bell-type correlations at an earlier time, which are encoded in the (non-vanishing) commutators. Moreover, because of causality, the commutators vanish at space-like separation. Therefore, the above difference is built up from interactions in the overlap between the past light-cones of the points $\x_1,\x_2$ and~$\x_3$. This implies that the information encoded in the correlators cannot be modified by local operations at late times. For quantum-vacuum fluctuations, the absence of a pole in the {\it folded} limit of the bispectrum thus becomes a unique signature of local causal evolution.

Limits of various $n$-point functions have been known to encode important physical information, e.g. \cite{Maldacena2,Creminelli,Assassi:2012zq,Flauger:2013hra,Goldberger:2013rsa,Arkani-Hamed:2015bza,Holman:2007na,LopezNacir:2011kk}. An enhanced `soft limit' (with soft internal or external momenta) is due to additional (light) fields, while the folded limit is enhanced for excited states. Yet, as we have demonstrated here,  the {\it absence} of an enhancement in folded configurations cannot occur with classical fluctuations, which would provide --- barring violations of locality --- striking evidence for the quantum origin of structure in the Universe. Although current observations are consistent with a Gaussian spectrum, surveys of increasing volume and sensitivity will continue the search for non-Gaussianity~\cite{Meerburg:2019qqi}. A true pole in the folded limit of the $n$-point functions would have diverging signal-to-noise~\cite{Babich:2004gb}.  While the non-zero width of physical particles will make the signal-to-noise finite, it is detectable nonetheless. However, enhanced dissipation may increase the width, thus reducing the signal-to-noise in the folded limit. At the same time, dissipation also increases the amplitude of the overall non-Gaussianty~\cite{LopezNacir:2011kk,Turiaci:2013dka,Porto:2014sea}, which is presently constrained by data~\cite{Akrami:2019izv}. Hence, as a matter of principle, it is possible (but potentially challenging) to distinguish between the spectrum of classical and quantum fluctuations. Moreover, as we have emphasized, the analytic structure of their respective shapes is clearly distinct, which suggests an analysis in position-space might provide a more stringent test, perhaps along the line discussed in~\cite{Munchmeyer:2019wlh}.

Even though we have restricted ourselves here to the case of density fluctuations, our results are also relevant for the quantum origin of primordial gravitational waves~\cite{Senatore:2011sp,Porto:2014sea,Shandera:2019ufi}. A detection would both be a measure of the energy scale of inflation and provide a putative signal of quantum gravity.  However, such a signal could also be the consequence of classical production mechanisms~\cite{Senatore:2011sp,Porto:2014sea}. Yet, as emphasized in \cite{Porto:2014sea,Mirbabayi:2014jqa}, classical production of gravitational waves also introduces a measurable non-Gaussian signature in the density perturbations. As we have shown, the {\it shape} of the associated non-Gaussianity will ultimately reveal the origin of these fluctuations.

Finally, the question we addressed here is  also of potential interest beyond cosmological applications.  We have shown that a local classical algorithm cannot reproduce the non-Gaussian correlations found from quantum evolution.  This is a limitation of classical computing.  It is unclear, on the other hand, if there exists a quantum algorithm that offers a significant computational advantage over classical approaches which include non-local (acausal) evolution. Nevertheless, it suggests an interesting connection between (quantum) cosmology and (quantum) computing, which deserves further study.\vskip 8pt
\noindent {\bf Acknowledgements}
We are grateful to Daniel Baumann, Roland de Putter, Raphael Flauger, John McGreevy, Alec Ridgway, Uro\v s Seljak, Benjamin Wallisch and Matias Zaldarriaga for helpful discussions. We would like to thank also the participants of the 
`Amplitudes meet Cosmology' workshop\footnote{\small \url{ https://www.simonsfoundation.org/event/amplitudes-meet-cosmology-2019/}} for useful conversations. D.\,G.~is supported by the US~Department of Energy under grant no.~DE-SC0019035. R.\,A.\,P. acknowledges financial support from the ERC Consolidator Grant ``Precision Gravity: From the LHC to LISA"  provided by the European Research Council (ERC) under the European Union's H2020 research and innovation programme (grant agreement No. 817791), as well as from the Deutsche Forschungsgemeinschaft (DFG, German Research Foundation) under Germany's Excellence Strategy (EXC 2121) `Quantum Universe' (390833306).

\clearpage
\appendix

\section{Bell-type Correlations}\label{app:comm}

For our example in the main text, with $H_{\rm int} = -\frac{\lambda}{3!} \dot \zeta^3$, the three-point function in the in-in formalism may be written as:
 \bea
&& \langle \zeta(\x_1,\tau)\zeta(\x_2,\tau) \zeta(\x_3,\tau)\rangle_q = -i\int_{-\infty}^{\tau} d\tau' d^3 \x' a^4(\tau') \Bigg[ \langle [H_{\rm int},\zeta(\x_1,\tau)] \zeta(\x_2,\tau) \zeta(\x_3,\tau) \rangle \nonumber  \\ && + \langle \zeta(\x_1,\tau)[H_{\rm int}(\x',\tau'),\zeta(\x_2,\tau)] \zeta(\x_3,\tau) \rangle + \langle \zeta(\x_1,\tau)\zeta(\x_2,\tau)[H_{\rm int}(\x',\tau'),\zeta(\x_3,\tau)]\rangle \Bigg] \nonumber \\
 && =  \lambda\int_{-\infty}^{t} dt' d^3 x' a^3(t') \Bigg[\dot G(t,t';\x_1-\x') \langle\dot\zeta^2(\x',t') \zeta(\x_2,t) \zeta(\x_3,t) \rangle  \\ && + \dot G(t,t';\x_2-\x')\langle \zeta(\x_1,t) \dot \zeta^2(\x',t') \zeta(\x_3,t) \rangle+ \dot G(t,t';\x_3-\x') \langle \zeta(\x_1,t)\zeta(\x_2,t)\dot \zeta^2(\x',t')\rangle \Bigg]\,, \nonumber
\eea
where in the second equality we restored the physical time, with $\dot G(t,t') 
=\partial_{t'} G(t,t')$, for notational convenience. As it turns out, the only difference from the classical calculation is the operator ordering. For Gaussian correlators, the quantum and classical power spectra are in fact related by
\beq
\langle \zeta(\x, \tau) \dot \zeta(\x',\tau') \rangle_c= \frac{1}{2} \left[ \langle \zeta(\x, \tau) \dot \zeta(\x',\tau') \rangle_q + \langle \dot \zeta(\x',\tau') \zeta(\x, \tau)  \rangle_q \right]\,,
\eeq
where, for the case of a de Sitter background, we have $\dot \zeta(\tau) =-H \tau \partial_{\tau}\zeta(\tau)$.
As a result, when computing correlation functions in perturbation theory we can substitute
\bea
 \langle \zeta(\x, \tau) \dot \zeta(\x',\tau') \rangle_q &=& \langle \zeta(x, \tau) \dot \zeta(\x',\tau') \rangle_c+ \frac{1}{2} \langle [\zeta(\x, \tau),\dot \zeta(\x',\tau') ]\rangle \\
 &=&\langle \zeta(\x, \tau) \dot \zeta(\x',\tau') \rangle_c +\frac{i}{2} H\tau'  \partial_{\tau'}G(\tau,\tau';\x_1-\x')\nonumber\,.
\eea
Hence, the difference between the quantum and classical three-point functions becomes 
\beq
\begin{aligned}
&\qquad\qquad\quad\langle \zeta(\x_1,\tau)\zeta(\x_2,\tau) \zeta(\x_3,\tau)\rangle_q -  \langle \zeta(\x_1,\tau)\zeta(\x_2,\tau) \zeta(\x_3,\tau)\rangle_c =\\ 
  &+  \frac{i  \lambda}{4} \int_{-\infty}^{\tau} d\tau' d^3 x' a^4(\tau') [\zeta(\x_1, \tau),\dot \zeta(\x', \tau')][\zeta(\x_2, \tau),\dot \zeta(\x', \tau')][\zeta(\x_3, \tau),\dot \zeta(\x', \tau')] \,,
\end{aligned} 
\eeq
where the commutators are $c$-numbers in this case. The above derivation can be extended to generic cubic interactions of the form  $H_{\rm int} = -\frac{\lambda}{3!}\Pi_\ell(\hat D_\ell \zeta )$, with $\hat D_{\ell=1,2,3}$ some local differential operators. In such cases, we find 
\begin{align}
 &\qquad \qquad \qquad\qquad\quad \zeta(\x_1,\tau)\zeta(\x_2,\tau) \zeta(\x_3,\tau)\rangle_q -  \langle \zeta(\x_1,\tau)\zeta(\x_2,\tau) \zeta(\x_3,\tau)\rangle_c = \\
 & \frac{i \lambda}{24}  \sum_\sigma \int_{-\infty}^{\tau} d\tau' d^3 \x' a^4(\tau') [\zeta(\x_1, \tau),\hat D_{\sigma(1)} \zeta(\x', \tau')][\zeta(\x_2, \tau),\hat D_{\sigma(2)} \zeta(\x', \tau')][\zeta(\x_3, \tau),\hat D_{\sigma(3)} \zeta(\x', \tau')]\,, \nonumber
\end{align}
where $\sigma$ is a permutation of $(1,2,3)$. This result can also be generalized to higher order interactions, with $\ell>3$. Although somewhat expected, this formula explicitly reveals how the quantum answer differs from the classical computation, by an integral over the product of three commutators. As we demonstrated here, this fact has deep consequences for the analytic properties of the correlation functions.  
\newpage
\section{Decay Width}\label{app:diss}
When particles are unstable, the Green's function develops new analytic structure. This can be illustrated simply for the harmonic oscillator. In the presence of a dissipative term we have
\beq
\ddot\zeta + \gamma \dot \zeta +\omega_0^2 \zeta = {\cal O}(t)\,,
\eeq
where ${\cal O}$ is the associated {\it noise}, expected from the fluctuation-dissipation theorem. Assuming strong dissipation, $\gamma \dot \zeta \gg \ddot \zeta$, the Green's function becomes,
\beq
G_\gamma(\omega) \simeq \frac{1}{\gamma} \frac{i}{\omega+i\Gamma} \quad \to \quad G_\gamma(t-t') \simeq \frac{1}{\gamma} e^{-\Gamma (t-t')}\theta(t-t')\,,
\eeq
with $\Gamma=\omega_0^2/\gamma$. The pole has moved to the complex plane, signaling an exponential decay in~time.

If the dissipation is strong enough, the power spectrum will be then dominated by the noise. This happens in an expanding universe, provided $\gamma \gg H$ \cite{LopezNacir:2011kk}. However, this is not the only change. The exact Green's function becomes more elaborate, see e.g. \cite{LopezNacir:2011kk}. In the limit $\tau \to 0$ and $\tau' \to \infty$, it takes the form
\beq
G_{\gamma}\left(z, z^{\prime}\right) \approx \frac{2^\nu \Gamma[\nu]}{\pi z} \frac{1}{z'^{\nu -1/2} } \sin( z')\,,\label{greenb}
\eeq
where $\nu = 3/2+ \gamma/(2H)$, $z = - k c_s \tau$ and $z' = -k c_s \tau'$, with $c_s$ the sound-speed. 
Notice the exponential decay in physical time has become polynomial decay in conformal time. 

In the strong dissipative regime, the non-Gaussianity is also dominated by the noise. Assuming the latter to be Gaussian, the former is produced by  non-linear effects. As shown in \cite{LopezNacir:2011kk}, the dominate term scales as $\gamma \k^2 \zeta^2$, generated by non-linear corrections to the response. The derivation is somewhat involved \cite{LopezNacir:2011kk}, yet the basic features in the bispectrum are captured by the~template 
\beq
\begin{aligned}
B(k_i,k_c) &\simeq \frac{\Delta_\zeta^4 f^{\rm eq}_{\rm NL}}{k_1^3k_2^3k_3^3}\Bigg[-k_1^3+k_2^3+k_3^3 +\frac{F(k_i,k_c)}{(-k_1+k_2+k_3+k_c)}  + \frac{G(k_i,k_c)}{(-k_1+k_2+k_3+k_c)^2}\Bigg]\,,\label{template}
\end{aligned}
\eeq
plus permutations, with $k_c \sim k_\star \log(\gamma/H)$ for some reference scale $k_\star$. The functions $F(k_i,k_c)$ and $G(k_i,k_c)$ are analytic in the momenta, and made out of products of the sort e.g. $k_1^2 k_2^2$ and $k_1^2k_2^3$, respectively, and  $f_{\rm NL}^{\rm eq}$ is the standard non-Gaussianity parameter in the equilateral configuration, e.g. \cite{Babich:2004gb}. (A similar template was introduced in \cite{Turiaci:2013dka} with $k_c/k_\star\simeq 3/4$ in the regime studied there.) The presence of the (shifted) poles is thus manifest.\footnote{We can also see the smoothing of the poles explicitly in the derivation of the three-point function. For example, consider the simple case of non-Gaussian noise with a non-zero three-point function. Then, in the folded limit,  $B(k_1,k_2,k_3) \sim \int_{-\infty} d\tilde\tau \tilde\tau^{4-3\left(1+\gamma/(2H)\right)}$, which is finite provided $\gamma/H>0$.}
There are, however, two main differences with respect to our flat-space intuition. First of all, the poles remain on the real axis, and secondly the width depends {\it logarithmically} on $\gamma/H$. This is due to the properties of the Green's function in~\eqref{greenb}. At the end of the day, the width smooths the would-be pole in the folded limit, producing instead an enhanced signal which is correlated to the size of the non-Gaussianity~\cite{LopezNacir:2011kk,Porto:2014sea}.

In principle, for sufficiently large $\gamma/H$, the pole may become `too wide' to be observed. While still classical, in practice this could be read as having no poles, thus misinterpreted as zero-point fluctuations. Yet, from the scaling with momenta in \eqref{template}, we see that with the poles removed the bispectrum does not reproduce the long-range correlations found in the vacuum. This can be seen already with the first term, which yields e.g. $1/(k_2^3k_3^3)$, that is clearly more {\it localized} in space than vacuum fluctuations.

\clearpage
\phantomsection
\addcontentsline{toc}{section}{References}
\bibliographystyle{utphys}
\bibliography{cosmoBell}

\end{document}